# Data mining and analysis of scientific research data records on Covid 19 mortality, immunity, and vaccine development- in the first wave of the Covid-19 pandemic

Petar Radanliev[1], David De Roure[1], Rob Walton[1]

Department of Engineering Sciences, University of Oxford[1]

Highlights

- We analysed scientific research records on Covid-19 with computable statistical methods.
- We present visualisations of interrelationships between scientific research data records on Covid-19.
- We compare the coverage of research on Covid 19 mortality, immunity, and vaccine development by organisations and countries.
- Despite political disagreements, countries (US and China) are still collaborating in Covid-19 research.

Abstract

Background and aims

Covid-19 is a global pandemic that requires a global and integrated response of all national medical and healthcare systems. Covid-19 exposed the need for timely response and data sharing on fast spreading global pandemics. In this study, we investigate the scientific research response from the early stages of the pandemic, and we review key findings on how the early warning systems developed in previous epidemics responded to contain the virus.

Methods

We conducted data mining of scientific literature records from the Web of Science Core Collection, using the topics Covid-19, mortality, immunity, and vaccine. The individual records are analysed in isolation, and the analysis is compared with records on all Covid-19 research topics combined. The data records are analysed with commutable statistical methods, including R Studio, Bibliometrix package, and the Web of Science data mining tool.

Results

[1] Corresponding author: Petar Radanliev, email: petar.radanliev@oerc.ox.ac.uk. ORCID: 0000-0001-5629-6857

Acknowledgements: Eternal gratitude to the Fulbright Visiting Scholar Project.

Funding: The authors acknowledge the UK EPSRC funding [grant number: EP/S035362/1] and the Cisco Research Centre [grant number 1525381].

From historical analysis of scientific data records on viruses, pandemics and mortality, we identified that Chinese universities have not been leading on these topics historically. However, during the early stages of the Covid-19 pandemic, the Chinese universities are strongly dominating the research on these topics. Despite the current political and trade disputes, we found strong collaboration in Covid-19 research between the US and China. From the analysis on Covid-19 and immunity, we wanted to identify the relationship between different risk factors discussed in the news media. We identified few different clusters, containing references to exercise, inflammation, smoking, obesity and many additional factors. From the analysis on Covid-19 and vaccine, we discovered that although the USA is leading in volume of scientific research on Covid-19 vaccine, the leading 3 research institutions (Fudan, Melbourne, Oxford) are not based in the USA. Hence, it is difficult to predict which country would be first to produce a Covid-19 vaccine.

Conclusions

We analysed the conceptual structure maps with factorial analysis and multiple correspondence analysis (MCA), and identified multiple relationships between keywords, synonyms and concepts, related to Covid-19 mortality, immunity, and vaccine development. We present integrated and corelated knowledge from 276 records on Covid-19 and mortality, 71 records on Covid-19 and immunity, and 189 records on Covid-19 vaccine.

## 1. Introduction

Since the emergence of Covid-19, the scientific literature on this pandemic has increased to a level that information can be overwhelming. While the cure is still nowhere in sight. In this study, we investigated three topics in relation to Covid-19, first mortality, second immunity, third vaccine. There is an argument that Covid-19 immunity can be established by surviving the infection, or through vaccine immunisation, both methods can apparently establish a heard immunity within a population [1]. However, we don't know for certain if heard immunity can be achieved with Covid-19. There is a suggestion that immune response could be similar to that of SARS and MERS [2]. One study suggests that the BCG vaccine used against tuberculosis, could reduce the effects of Covid-19 [3]. Another study argues that the BCG vaccine may not reduce the Covid-19 mortality [4].

Given that the most severe inflammation in the lungs is directly caused by the immune response, some articles even argue that 'good general health may not be advantageous' when patients have advanced to the severe stage [5]. In addition, studies on antibody-dependent enhancement are conflicting, because antibodies 'can enhance virus uptake by cells' [6], but also 'may also inhibit virus entry' [7]. One study argues that physical activity and exercise can improve immune response to Covid-19, and '.. prevent viruses and other pathogens from gaining a foothold.' [8]. One study even argues that obese patients are more contagious than lean patients, attributing obesity to spread of contagion [9]. The same study also argues that losing weight, and moderate exercise can improve the immune response to Covid-19.

This is just the beginning of the rising scientific records and information available on Covid-19. Despite the apparent conflict in information, this is a natural process in the early stages of searching for a solution to a fast spreading global pandemic. One things is certain, that is the need to invent a vaccine that can be manufactured and distributed to immunise the entire global community from Covid-19 [10]. But even if such vaccine was developed, it would still take between 3-10 months for commercialisation [11]. Alternatively, for heard immunity, we would need to survive a number of repeated waves of Covid-19, until 60 to 70% of the population can develop immunity [12]. These are difficult choices, and we can benefit from understanding all the scientific advice, which currently stands at 5,208 research papers on the Web of Science Core Collection alone.

In this study, we present a computable statistical analysis of the current scientific research records, and we present conceptual maps that integrate and corelate the knowledge of current studies on Covid-19 mortality, immunity, and vaccine development.

## 2. Methodology

In this article, we used computable statistical methods for data mining and analysis on Covid 19 scientific literature. The specific focus of the data mining and analysis was on three specific topics: (1) Covid 19 mortality; (2) Covid 19 vaccine; (3) Covid 19 immunity. We performed three different searches on the Web of Science Core Collection[2], one on each topic. Then we analysed the data records with R Studio. We used the 'biliometrix' package for the statistical analysis [13]. In our data mining for scientific data records, we used the Web of Science Core Collection. The Web of Science Core Collection contains over 21,100 peer-reviewed, high-quality scholarly journals published worldwide (including Open Access journals) in over 250 sciences, social sciences, and arts & humanities disciplines. For disclosure, we only used the Web of Science Core Collection database, we didn't download data records from PubMed, or Google Scholar. We hope that other researchers can do that, and compare the results. Journals have a limit on figures, and we exceeded that limit in this article. Analysing different data sets, would produce even more figures. Our rationale was that, since the Web of Science Core Collection contains most accredited journals, we didn't see how the results would be different today. It could be more useful to run that analysis at a later stage of the pandemic, then compare the visualisations.

## 3. Bibliometric analysis

Bibliometrics in this article refers to the use of computable statistical methods, to:

a) analyse scientific research records and identify research relationships in journal citations;
b) to quantitatively assess the most dominant keywords;
c) to identify the interrelationship between the problems investigated by different organisations and countries;
d) to compare the coverage of research topics by countries;
e) to create list keywords in groups of synonyms and related concepts (e.g. Covid 19 thesauri); and
f) to measure term (keywords) frequencies.

The bibliometric analysis in this article differentiate from scientometrics. Although both research fields study, measure and analyse scientific literature, scientometrics is used to measure the impact of research papers and journals, to understand scientific citations, usually as measurements in policy and management. Instead, our bibliometric analysis is focused on creating conceptual map (of keywords, synonyms and related concepts) through factorial analysis, creating collaboration network map (of countries or organisations), and categorising the keywords, synonyms and related concepts, then relating these groupings to countries or organisations, e.g. in a three-fields plot.

---

[2]http://apps.webofknowledge.com/WOS_GeneralSearch_input.do?product=WOS&search_mode=GeneralSearch&SID=C3DVq2qEsnSXLxyxR1u&preferencesSaved=

## 4. Data mining

We used the Web of Science Analyse Results tool to get a fast understanding of the scientific literature data records on Covid-19. For this visualisation, we wanted to analyse all data records on Covid -19. Hence, our data mining included all data records on Covid-19, without using Booleans (e.g. AND, OR, NOT, SAME, NEAR).

### 4.1. Rationale for data mining for records periodically - creating snapshots in time

In the Web of Science Analyse Results data mining on the topic of Covid-19, we identified 5,210 (on date 01 June 2020). The search identified 5,208 results produced in 2020, and 2 data records from 2019. We checked the 2 data records, and the published data showed was October 2019. This was puzzling, because we didn't know about Covid-19 until later date, so we reviewed the two papers manually, and we found that the actual publication date was in March and April 2020. This reemphasised our argument that we should perform regular analysis of data records, because even the most established databases, depend on the journals data. If that data is incorrectly structured, or at least not structured to the requirements of the data base, we could get incorrect reading. In this article, we are creating a snapshot in time, and this is a more reliable representation than searching for data records from specific months at a later stage. For example, if we conducted this data mining prior to March, the two data records from March and April would not have been included, because these data records didn't exist until March and April - respectively. Adding to this argument, the Web of Science databases are separating data records by year, not by months. In our search, we could not separate data records by months, hence, if this data mining is performed in future years, it would be even more challenging to conduct factorial analysis, three-field plots, and conceptual maps on the scientific data records for this period in time. The results would present all data records from 2020, and we are interested in the analysis of these data records from the first wave, and the early stages of the pandemic. Since the journal editorial and peer-review process last from 3-6 months, we can assume that the records we are analysing today, are submitted in the early stages of the pandemic. Hence, we can analyse the keywords, synonyms and related concepts from the first wave of the Covid-19 pandemic.

### 4.2. Data mining on the topic of Covid-19

In our first visualisation, we used the Web of Science Analyse Results tool. We wanted to identify the funding agencies that supported most scientific research records in the first wave and the early stages of the Covid-19 pandemic (see Figure 1).

*Figure 1: Data mining on the topic of Covid-19: Tree-map of funding agencies that supported most scientific research records*

From Figure 1, we can see that when the scientific data research is separated in organisations, Chinese organisation emerges as a leader from this data mining visualisation. We can also see that in Figure 1, there are multiple organisations from individual countries. In our initial community discussions, we noticed some negative comments, stating that we have '..purposely split the two largest US institutions to halve their contribution levels'. We wanted to re-emphasise that one country, can have multiple institutions. In Figure 1, we analysed the data records by organisation, not by nation. In the interest of eliminating such negative comments, we analysed the data records by country in Figure 2.

*Figure 2: Data mining on the topic of Covid-19: Bar-graph of countries that supported most scientific research records*

From the bar-graph in Figure 2, we can see that USA has produced most scientific research. From the bar-graph in Figure 2, we notice a significant increase in data records originating from the USA since our earlier study [14]. It would have been interesting to conduct such analysis at much earlier stages of the pandemic. Since from the bar-graph in Figure 2, we can see that countries that were most affected (USA, China, England, Italy), produced most data records, and we have noticed a significant increase from the USA - since our previous study, we can just wander if the efforts of the scientific community increased as the pandemic was increasingly more present in those countries. This could signify that the scientific community didn't act, and ignored the warning signs, until they were faced with the tragedy. However, without conducting a separate data mining of records, specifically from that early stage (January, February, and March), this would remain an open question.

## 5. Data analysis

In the data mining for analysing the data records with R Studio, we used more specific search. This resulted with far less data records, because in this data mining effort, we used Booleans to identify records on the three topics we investigated (mortality; vaccine; and immunity).

### 5.1. Analysis of scientific data records on Covid-19 and mortality

From our search on TOPIC: (covid 19) AND TOPIC: (mortality), we identified 276 data records. Similarly, to the earlier visualisations, we used the Web of Science Analyse Results tool (Figure 3 and Figure 4).

*Figure 3: Data mining on the topic of Covid-19 and mortality: Tree-map of Universities that produced most scientific research records on Covid-19 and mortality*

From the tree-map in Figure 3, we can see that Chinese universities are leading the research effort on the topics of Covid-19 and mortality. It would be interesting to compare these results after a period of time, because in our recent study on this topic, we identified through historical analysis of research studies on pandemics and epidemics, that Chinese universities have not been leading on these topics [14]. Nevertheless, we can clearly see from Figure 3 that Chinese universities are strongly dominating the research on these topics. To compliment this analysis, we compare the tree-map from Figure 3, with a new tree-map in Figure 4, which categorises data records by country.

*Figure 4: Data mining on the topic of Covid-19 and mortality: Tree-map of countries that produced most scientific research records on Covid-19 and mortality*

In Figure 4, we can see that the most affected countries (USA, China, England and Italy) are leading the research efforts on Covid-19 and mortality. At this stage, we wanted to investigate these data records further, and we faced limitations in the capabilities of the Web of Science Analyse Results tool. We used the same data record, but with the R Studio – in Figure 5.

*Figure 5: R Studio - Three-fields plot: left – keywords from the data records, middle – countries, right – authors affiliations.*

We used the data records in R Studio, and we created a three-field plot in Figure 5, separating keywords of the research studies by country. From the Figure 5, we can identify which university, is researching specific topics related to Covid-19 and mortality. We wanted to visualise which countries collaborate most. To visualise the collaborative relationships, we plotted this data file in a country collaboration map - in Figure 6.

*Figure 6: R Studio – research data records on covid-19 and mortality, country collaboration map*

From the country collaboration map in Figure 6, we can see a strong collaboration line between the US and China and Italy, and strong research relationship between China and UK. But surprisingly, this analysis shows that UK is not collaborating with the US and Italy as strongly as with China. We wanted to evaluate this result further, before making any conclusions. To investigate this further, we created a social structure – collaboration network, with specific countries in the network parameters in Figure 7.

*Figure 7: R Studio – collaboration network on covid-19 and mortality, with specific countries in the network parameters*

By analysing the specific countries of interest, the collaboration network in Figure 7 is more detailed, and we can see two different clusters (in green and purple). Although we identified these two clusters, we cannot analyse these collaborations in more detail with the social structure of this collaboration network. For our final analysis of the data records on Covid-19 and mortality, we wanted to identify the related topics, the keywords, synonyms and how these concepts are related. For this, we used factorial analysis to create a conceptual structure map, with multiple correspondence analysis (MCA) – in Figure 8.

*Figure 8: R Studio – factorial analysis, conceptual structure map with multiple correspondence analysis (MCA)*

From the factorial analysis in Figure 8, we can see how different concepts are corelated in research studies. Worth mentioning that any computable statistical software, would scan for keywords in the data records, and apply the factorial analysis to identify multiple correspondence analysis (MCA) to all keywords. Occasionally, such keywords could be irrelevant, for example, it seems that from all the data records, the keywords Wuhan and China were present in the data records. Hence, the statistical software extracted these keywords and placed in the factorial analysis. This is not deliberate, it's just how the statistical software extracts keywords. But it is strange that we don't see other countries and regions in the conceptual structure map. Could this simply be because of research papers including Wuhan, and China in the introduction of the paper? Or could it mean that these studies are using data records shared by medical institutions in Wuhan and China? Without a further analysis, this is difficult to know. But since this is not the research objective of this article, we conclude the data analysis on Covid-19 and mortality with the conceptual structure map.

### 5.2. Analysis of scientific data records on Covid-19 and immunity

The search for data records on Covid-19 and immunity produced only 71 results. The first thing we wanted to identify from these data records, was the Covid-19 and immunity related fields in Figure 9.

*Figure 9: Tree-map of research areas identified from the Covid-19 and immunity scientific data records*

In the first wave, we have seen many discussions in the media on heard immunity, Covid-19 risk factors, such as obesity, and the value of daily exercise. We wanted to identify if there is a relationship between any of these topics in the scientific literature. The record count in each area is the total number of articles published. But from categorising the data records in Figure 9, we cannot identify the content of the articles. Therefore, we used the R Studio to analyse this data record further. We used the R Studio to design a co-occurrence network of the keywords from the data records. We didn't use the articles keywords, because the visualisation was generic and didn't analyse the text, similarly to the Figure 9. Instead, we used the most occurring keywords from the data records text, and we designed the co-occurrence network as a sphere - in Figure 10

*Figure 10: R Studio – Covid-19 and immunity, co-occurrence network of the keywords extracted from the data records*

To build the co-occurrence network as a sphere (in Figure 10), we applied equivalence normalisation, with a sphere as the network layout, and Louvain clustering algorithm. This was our attempt to visualise the relationships, with colour coding the keywords, synonyms and related concepts. Although we can see from Figure 10 how these keywords, synonyms and related concepts are investigated in the context of Covid-19 and immunity, we don't see a clear direction in this research field. To analyse the data record further, we applied factorial analysis on this data record as well, but this time we build the conceptual structure map with the multidimensional scaling (MDS) method – in Figure 11.

*Figure 11: R Studio – factorial analysis on Covid-19 and immunity, conceptual structure map with the multidimensional scaling (MDS) method*

In the Figure 11 - conceptual structure map, we can see few different clusters appearing, and the main cluster, contains references to exercise, inflammation, smoking, obesity and many additional factors. The multidimensional scaling (MDS) method, enables this visualisation of key concepts related to Covid-19 and immunity. These concepts, are extracted from all of the 71 scientific research studies that we found on Covid-19 and immunity. Since the factorial analysis in Figure 11 is designed from the keywords found in the text of the data records, and not from the keywords provided by authors, we can expect some irrelevant concepts. Again, this is simply how the statistical software works, it extracts all keywords that are found to be repeating in different data records.

### 5.3. Analysis of scientific data records on Covid-19 and vaccine

The final data record we analysed was on Covid-19 and vaccine. In the data mining for records, we searched the TOPIC: (covid 19) AND TOPIC: (vaccine) and we identified 189 records. As with the previous data records, firstly we analysed the data records by organisations (in Figure 12) and countries (in Figure 13).

*Figure 12: Bar-graph of the leading organisations in the scientific research on Covid-19 vaccine – designed with the Web of Science data mining tool*

What we can see from the Figure 12, is that the leading research organisations on Covid-19 vaccine are not based in the leading country on Covid-19 vaccine - in Figure 13. Fudan University in Shanghai, University of Melbourne and University of Oxford are the top 3 institutions, with the same output (Figure 12). While in Figure 13, we can see that the USA is the overall leader in research on Covid-19 vaccine. This makes it even harder to predict which country would be first to produce a Covid-19 vaccine.

*Figure 13: Tree-map of the leading countries in the scientific research on Covid-19 vaccine – designed with the Web of Science data mining tool*

From the Figure 13, it seems that the USA which is at present the number one in total research output, is on track to produce a vaccine first – based on the most research output. But if we consider that to produce a new vaccine is a lengthy process, then one could argue that countries and organisations that started earliest, would be best placed to produce a Covid-19 vaccine. This is impossible to predict, but it would be interesting to compare these results over time. This analysis and the visualisations can preserve the present understanding and efforts, and be used as a snapshot in time by future researchers, analysing the Covid-19 research records. Since we cannot answer these questions from the current data records, we applied computable statistical analysis to look for further insights on the relationships between organisations, countries and Covid-19 vaccine. In Figure 14, we designed a three-fields plot, with countries on the left, keywords from the data records in the middle, and universities on the right.

*Figure 14: three-fields plot: countries - left, keywords - middle, universities - right.*

In this three-fields plot (Figure 14), we wanted to identify the relationships between the research findings from the leading organisations (in Figure 12) and compare with the overall national research efforts (in Figure 13). To design the three-fields plot (in Figure 14), we extracted the keywords from all the data records, and we associated the keywords with countries, and organisations. What becomes visible from the Figure 14, is the lower research output of the US in the keywords that are most present in all data records. Since the US is the leader in the overall research on Covid-19 vaccine at present, we wanted to determine if the US research is focused on different research areas, and not related to the

keywords that are taken as most represented in the combined research records from all countries. We continued the analysis with a topic dendrogram (in Figure 15).

*Figure 15: topic dendrogram on Covid-19 and vaccine keywords*

To design the topic dendrogram in Figure 15, we used factorial analysis with multidimensional scaling as the method for parameter analysis of the different keywords. Although we can see how clusters develop from the correlations of the keywords, it is unclear which countries and organisations collaborate to create these clusters. The final data analysis we performed on this data record, was a design of social structure as a collaboration network of the current research efforts on Covid-19 and vaccine development (in Figure 16).

*Figure 16: Covid-19 vaccine social structure of the research collaboration network*

From the collaboration network in Figure 16, we can see that despite the current political issues between the USA and China, the scientific research on Covid-19 vaccine is very strong. This bring some optimism in the search for a Covid-19 vaccine. This also supports the previous findings in Figure 7, on the USA and China collaboration network on covid-19 and mortality research. It seems that despite the lack of collaboration between the leading countries on Covid-19 research (see Figure 6), the scientific research is ongoing, but possibly the visibility (of these research collaborations) is limited.

## 6. Discussion

The findings of this study confirm that despite the political disagreements, the collaborations on Covid-19 scientific research, between the countries leading this research, is strong. The study also finds some correlation between the countries worst affected by Covid-19, and countries most productive in Covid-19 research. We can see the same countries showing up in all of the visualisations as leaders. These same countries are also the worst affected by Covid-19. With some exceptions (e.g. Germany), majority of countries that got affected, produced the most output. There is a second correlation between the data records, that's is, countries tend to produce more output as they get affected by Covid-19. For example, organisations in China are leading in the research efforts in the early stages of the pandemic. But as Covid-19 spread to other countries, and reached the UK, USA, Italy, India, the output of these countries increased. We can expect to see this analysis changing fast, and if we repeat the same analysis after some time, we can expect different organisations to be in the lead. The **implications for future research** from this study is the visualisation of the scientific data records, as on the date (01 June 2020) the records were collected. Future research can use these visualisations, and replicate the analysis, with additional data, as such data becomes available. This study presents a snapshot in time, which will preserve the state of Covid-19 research as of 01 June 2020. **Future research dimensions** should include firstly a comparison of the data records, to seek changes in the topography of the data records. Secondly, future research can use the visualisations and the data records from this study, to analyse the historical response from the first wave of Covid-19. The implications for organisations and practitioners is a clear comparison of output, so different organisations can visualise and compare their performance in the first wave of

Covid-19. Evaluation of past performance could be used in improving the response in the second wave, or in future pandemics. For example, organisations that reacted slower, can develop new and improved response strategies. The study identified the best performs, and practitioners should learn from these organisations. We really need to learn to start picking up the phone and communicating with other organisations, and learning about the pandemic. In terms of **using this research for improving medical systems**, we could refer to the factorial analysis, where we extracted and corelated keywords from all data records, to create conceptual maps of how keywords, synonyms and concepts are related. While we hear a lot of different risk factors (e.g. obesity, smoking, etc), we usually see these risk factors in isolation. From the factorial analysis in this study, we can visualise all the risk factors in clusters. Finally, we include a separate section on confirming validity and eliminating bias.

### 6.1. Eliminating bias and confirming validity

We considered the limitations of qualitative literature review, especially the limitations in value caused by bias in collecting data records manually. Data records can be selected to represent a biased viewpoint (e.g. one nation, organisation, or author - trying to show a better performance than others). We used statistical data mining approach. The records we collected from the Web of Science Core Collection, include all data records (as found on the date 01 June 2020) on the three topics we analysed (Covid 19 - mortality; vaccine; and immunity). This eliminated bias in selecting the data records. Since we used well know, and established computable statistical programs, the risk of incorrect representation of sources, and bias in the analysis, was eliminated by the statistical software. In the spirit of reproducible research, we include our data records in the submission of this article. In brief, to eliminate bias, we used data mining approach. To confirm validity of the results, we performed a diverse set of computable statistical analysis.

## 7. Conclusion

In this research study, we conducted data mining with the Web of Science Analyse Results tool and identified 5,210 records on Covid-19. We created data visualisations to identify the countries, institutions and organisations leading the scientific research on Covid-19. In the original visualisation of the data records, Chinese organisations emerge as leaders. By the time we finished the paper, we conducted a second data mining and visualisation. From the second visualisation, we noticed a significant increase in data records originating from the USA. This could represent a higher research output when the effect of the pandemic is higher, but it is hard to determine this with certainty, without additional analysis of future data records. Hence, this study presents a snapshot in time, and can be used by future studies to investigate the relationship between research output and spread of the pandemic.

We focused our analysis of scientific research records on identifying research relationships in journal citations, and we confirmed that despite political and trade disputes, there is a strong collaboration on Covid-19 research between the most affected countries. We used the R Studio to quantitatively assess the most dominant keywords, and we present visualisations that integrate the keywords from all scientific research studies. With the overwhelming number of studies emerging every day, we identified a process of visualising the keywords from all studies. We used the statistical software to measure the keywords frequencies and we presented conceptual maps based on the statistical analysis. In the data visualisations, we also presented the keywords in groups of related concepts and we identify the interrelationship between the problems investigated by different organisations and countries. We present the keywords in a dendrogram, presenting different clusters of Covid-19 research and we designed three-field plots, connecting different Covid-19 topics to countries and organisations. Apart from identifying existing collaborations, the data visualisations enable quick identification of related research in different organisations and institutions. This should promote an increased collaboration between organisations and institutions conducting research on similar or related topics, and speed up the research process/output. We also compare the coverage of research topics by countries, and we present visualisation in the form of collaboration maps, structured around the 3 different topics we investigated. We used tree-maps and bar-graphs to analyse the data records, in three separate chapters, corresponding to Covid 19: mortality, immunity, and vaccine. Then we designed three-fields plots, and country collaboration maps, to analyse the three topics further and we used multiple correspondence analysis (MCA) to create conceptual structure maps of the collaboration networks. The data visualisations presented in this research, can be used to as a snapshot in time by future research studies on this topic. At present, the visualisations can be used to review the state of Covid-19 research, and to navigate through the increasing volume of scientific publications on the state of the pandemic.

### 7.1. Limitations

This research study is based on the limited data records available on Covid-19 and the topics investigated. Future research should use the findings as a record of the scientific data records in this time period. Research data records are changing every day, and this study wanted to present a snapshot in time, to assist future analysis of the Covid-19 response in the first wave of the pandemic.

## 9. Figures:

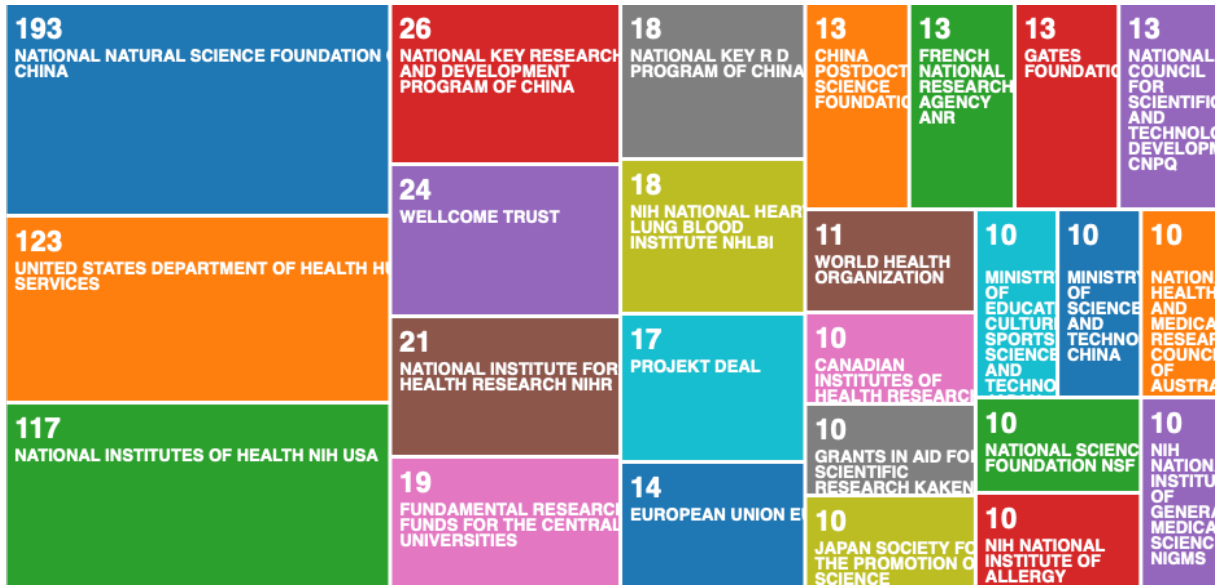

*Figure 17: Data mining on the topic of Covid-19: Tree-map of funding agencies that supported most scientific research records*

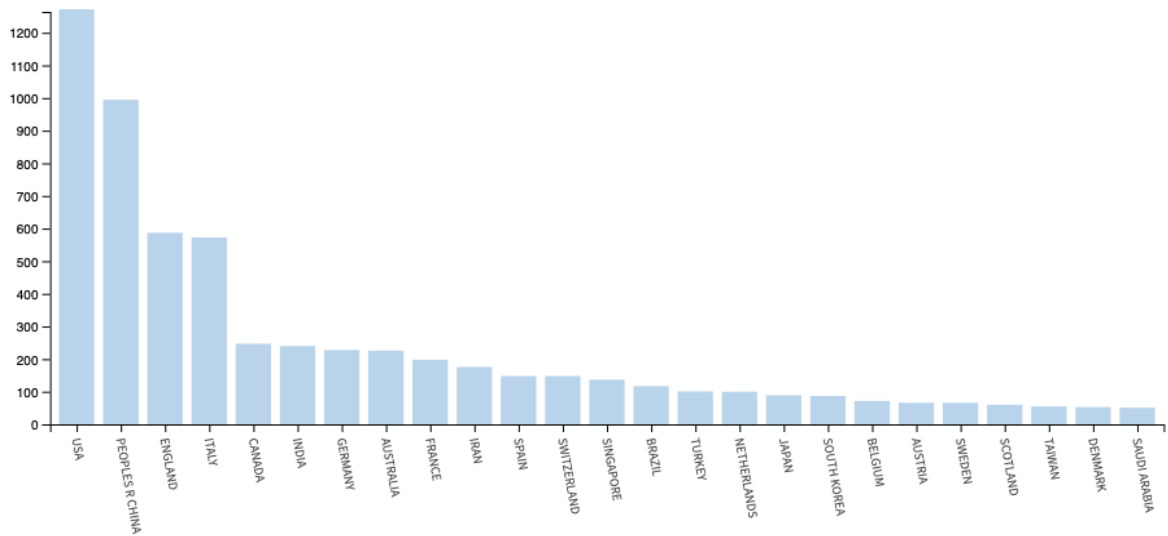

*Figure 18: Data mining on the topic of Covid-19: Bar-graph of countries that supported most scientific research records*

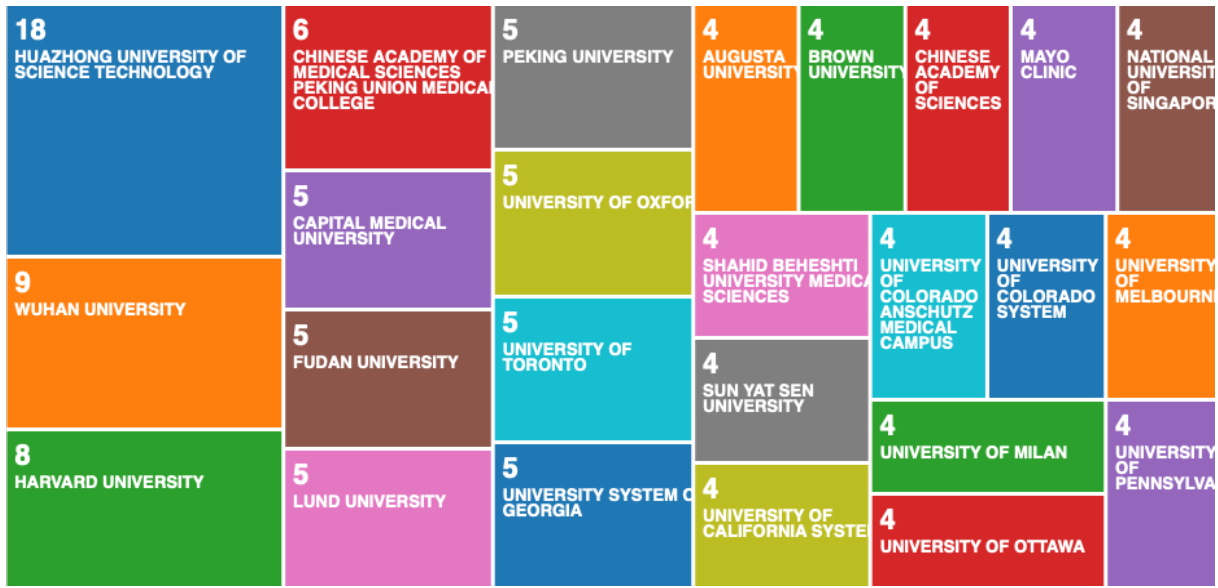

*Figure 19: Data mining on the topic of Covid-19 and mortality: Tree-map of Universities that produced most scientific research records on Covid-19 and mortality*

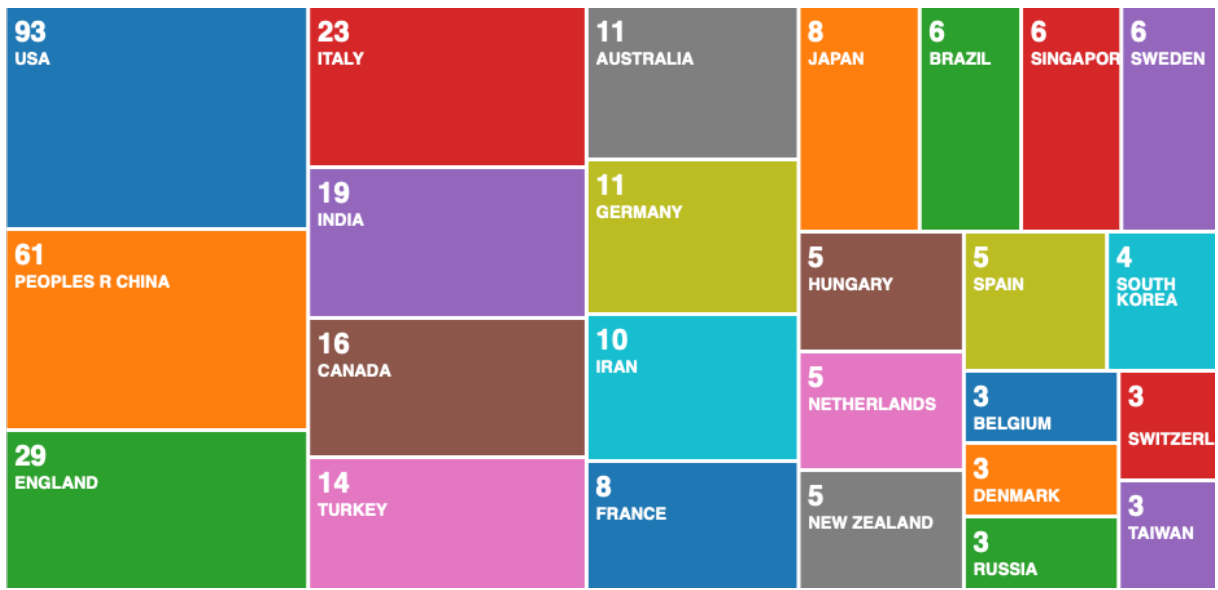

*Figure 20: Data mining on the topic of Covid-19 and mortality: Tree-map of countries that produced most scientific research records on Covid-19 and mortality*

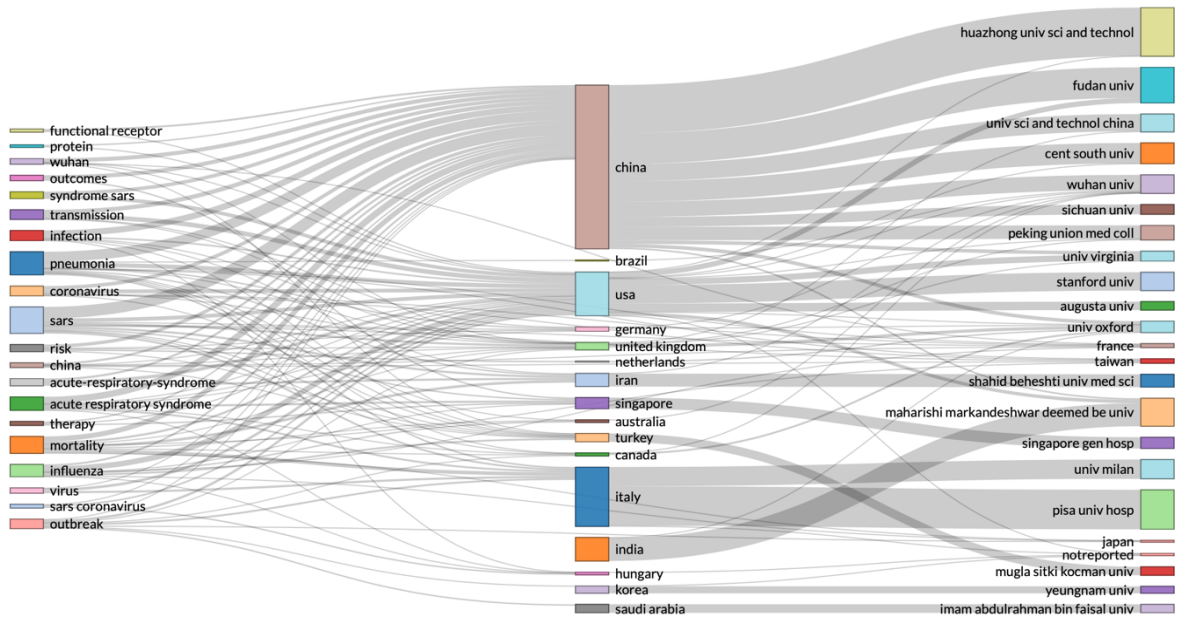

*Figure 21: R Studio - Three-fields plot: left – keywords from the data records, middle – countries, right – authors affiliations.*

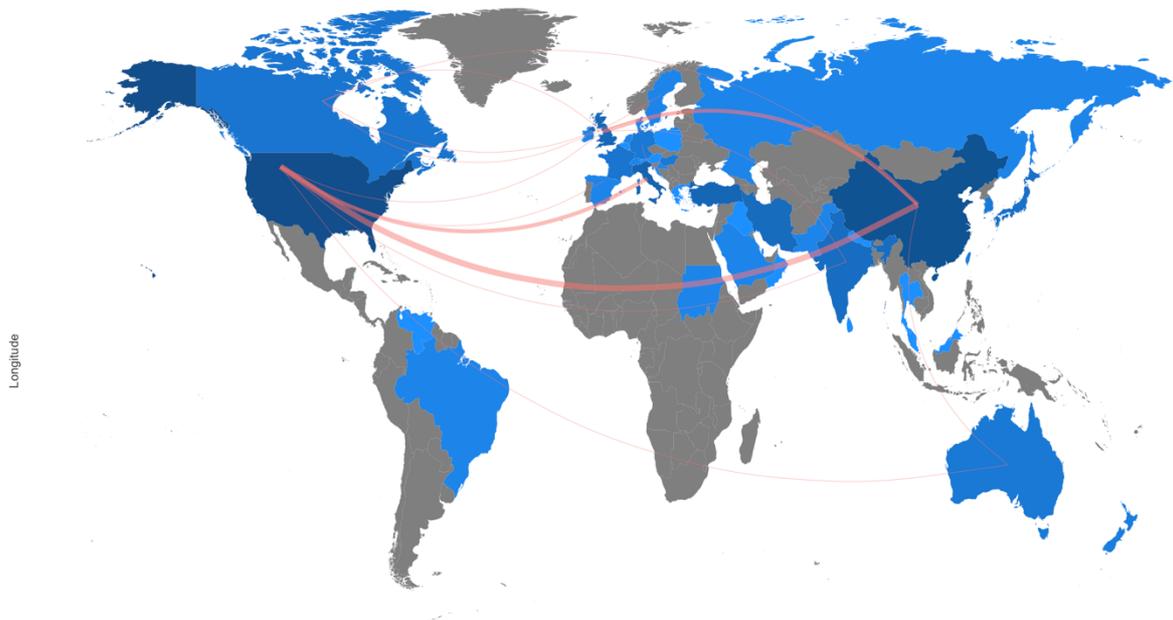

*Figure 22: R Studio – research data records on covid-19 and mortality, country collaboration map*

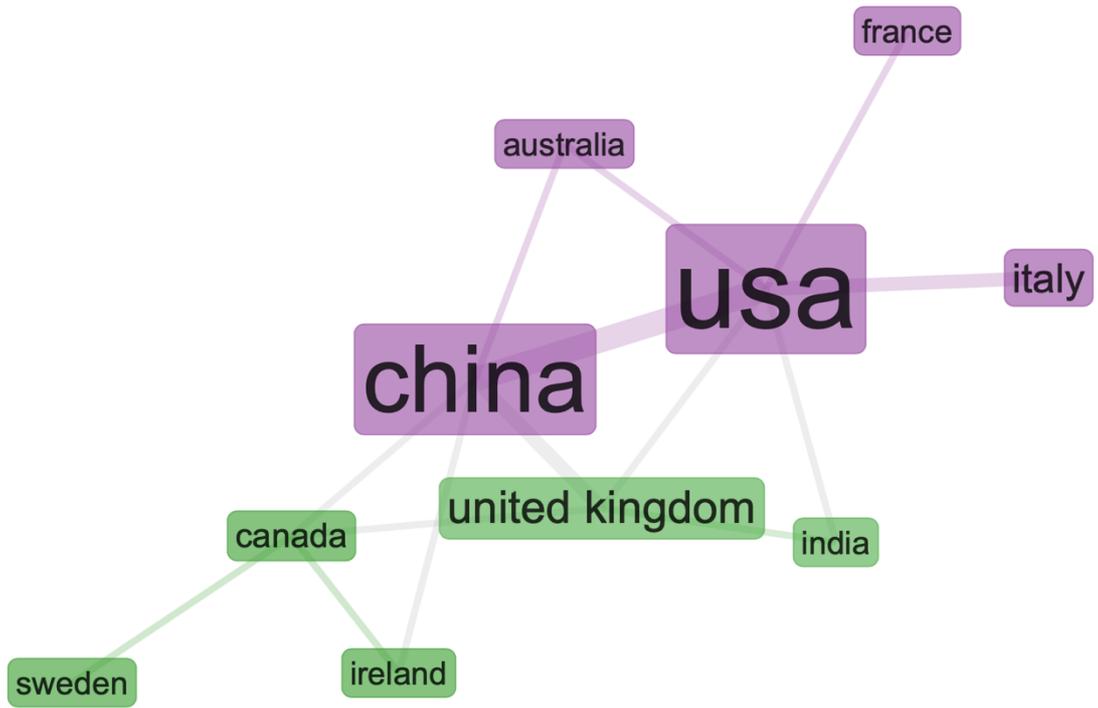

*Figure 23: R Studio – collaboration network on covid-19 and mortality, with specific countries in the network parameters*

*Figure 24: R Studio – factorial analysis, conceptual structure map with multiple correspondence analysis (MCA)*

*Figure 25: Tree-map of research areas identified from the Covid-19 and immunity scientific data records*

*Figure 26: R Studio – Covid-19 and immunity, co-occurrence network of the keywords extracted from the data records*

*Figure 27: R Studio – factorial analysis on Covid-19 and immunity, conceptual structure map with the multidimensional scaling (MDS) method*

*Figure 28: Bar-graph of the leading organisations in the scientific research on Covid-19 vaccine – designed with the Web of Science data mining tool*

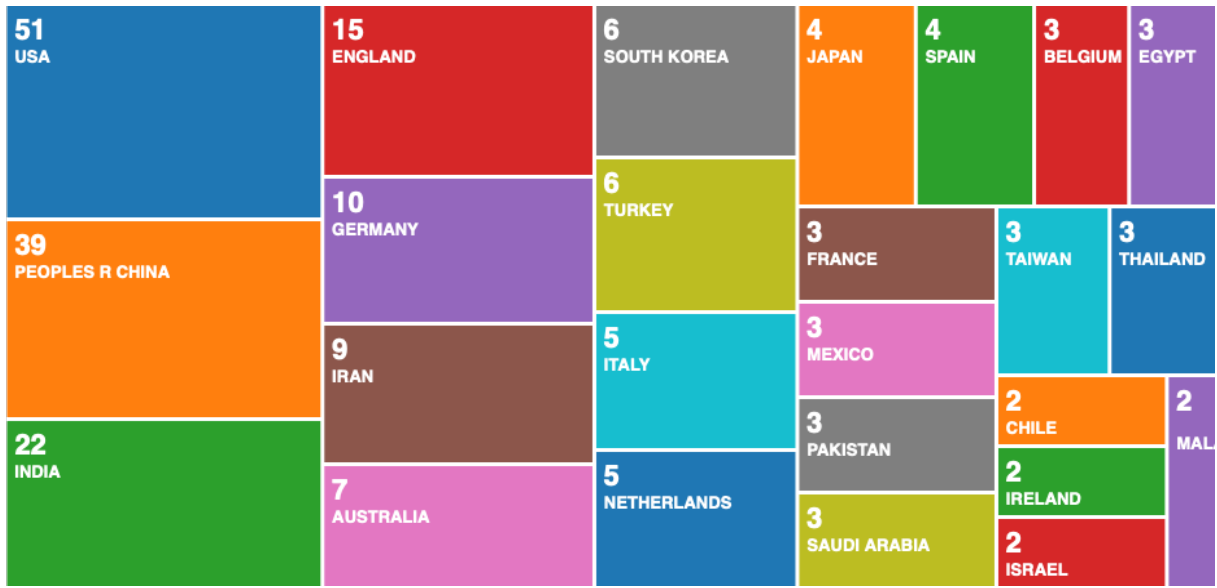

*Figure 29: Tree-map of the leading countries in the scientific research on Covid-19 vaccine – designed with the Web of Science data mining tool*

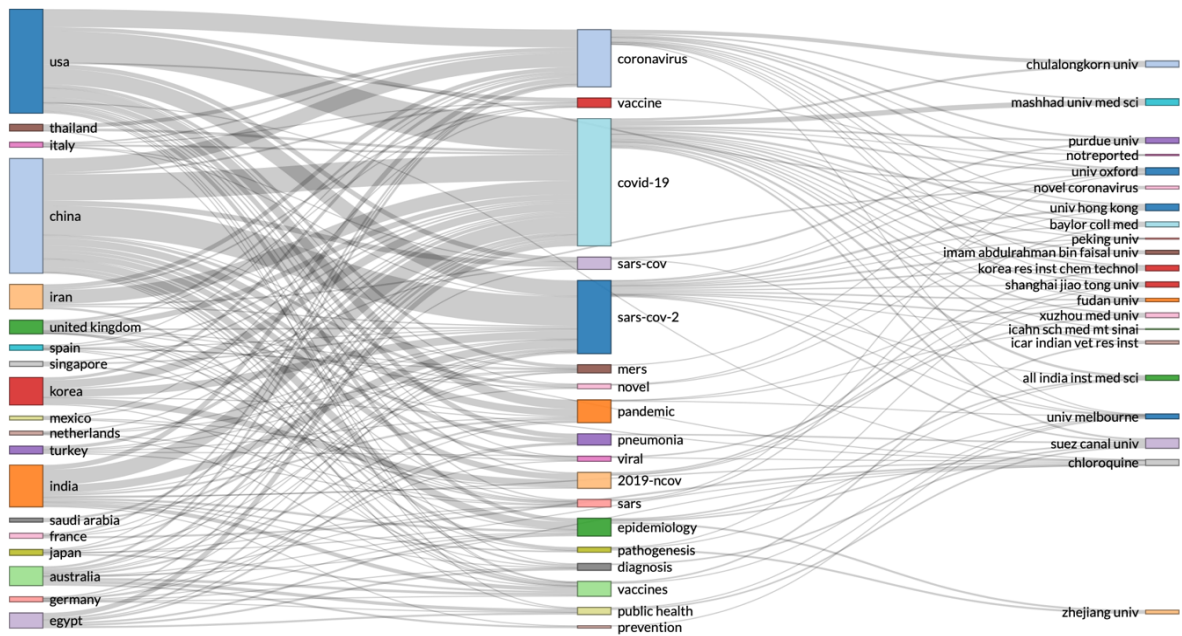

*Figure 30: three-fields plot: countries - left, keywords - middle, universities - right.*

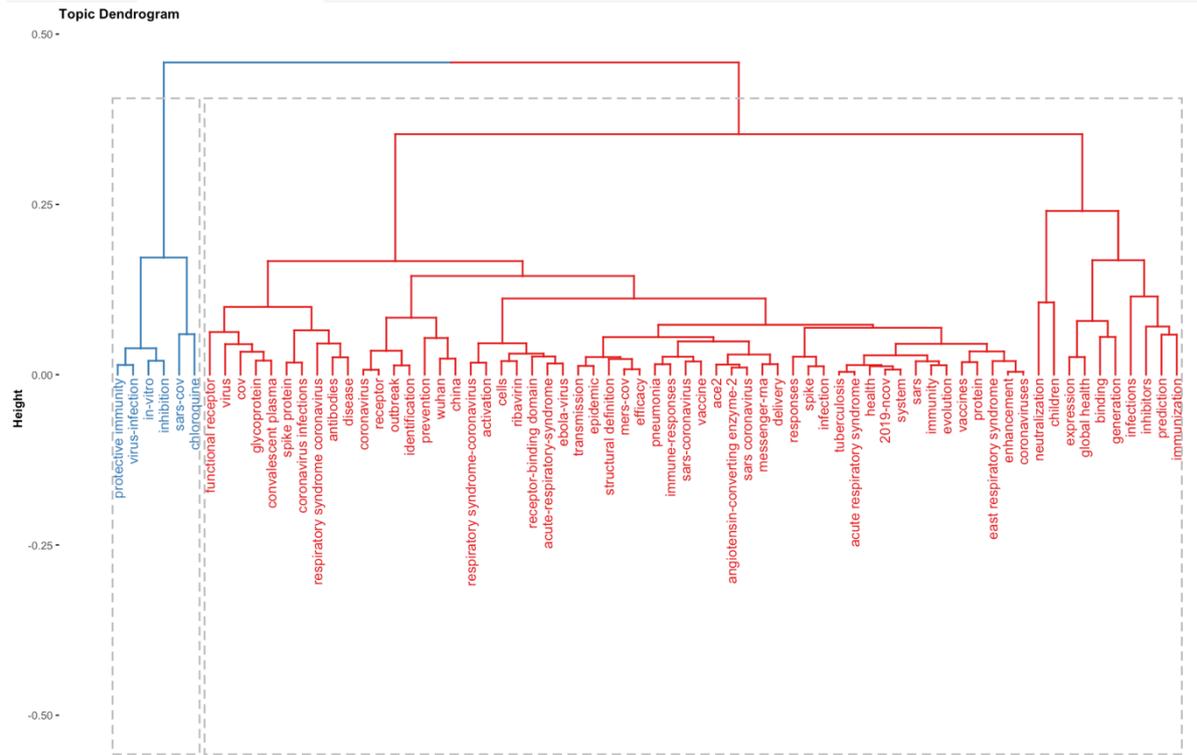

*Figure 31: topic dendrogram on Covid-19 and vaccine keywords*

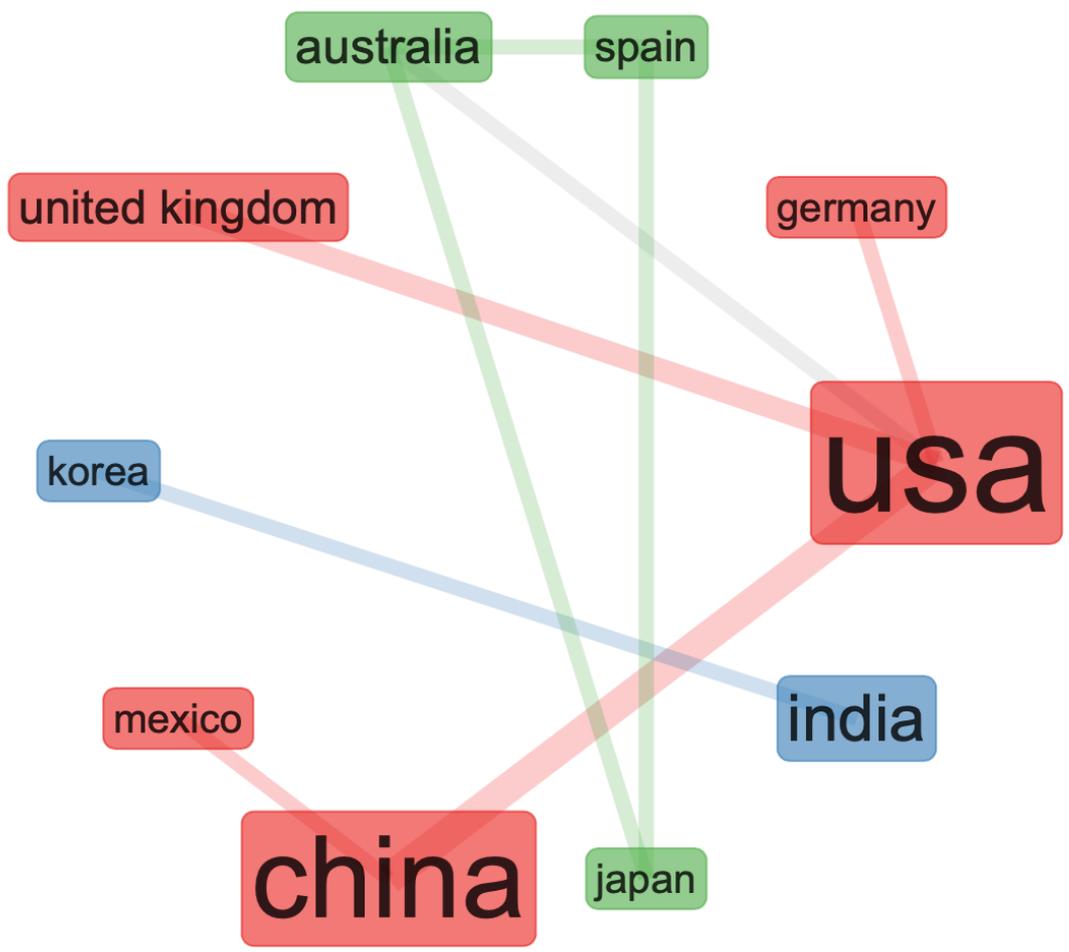

*Figure 32: Covid-19 vaccine social structure of the research collaboration network*